\documentclass[floatfix, twoside, twocolumn, aps]{revtex4}
\usepackage{epsfig}
\begin{document}

\title{Large stock price changes: volume or liquidity?}

\author{Philipp Weber and Bernd Rosenow}

\affiliation{Institut f\"ur Theoretische Physik, Universit\"at zu
K\"oln, D-50923 Germany}

\date{January 8, 2004}

\begin{abstract}
  We analyze large stock price changes of more than five standard
  deviations for i) TAQ data for the year 1997 and ii) order book data
  from the Island ECN for the year 2002.  We argue that large price
  changes are not due to large trading volumes.  Instead, we find that
  extreme price fluctuations are mainly caused by a  low density of
  limit orders stored in the order book, i.e.  a small liquidity.

\end{abstract}

\maketitle

The discovery of power law distributions for commodity
\cite{Mandelbrot63} and stock price changes \cite{Lux96,Go+98}
together with the relevance of this discovery for the practical
problem of risk management has spurred a large amount of interest in
the price process in financial markets. For physicists, the power law
distribution of price changes is very appealing as the appearance of a
power law is reminiscent of universality and critical phenomena, thus
suggesting that there might be a basic and universal mechanism behind
the distribution of price changes. Many phenomenological as well as
microscopic models have been developed, which are able to explain the
main stylized facts about financial time series \cite{Takayasu02}.

In order to understand the mechanism underlying the empirically
observed return distribution in detail, one needs to study the price
impact of trades. Besides the influence of news breaking, stock prices
change if there is an imbalance between supply and demand. If more
people want to buy than to sell, stock prices will move up, if more
people want to sell than to buy, they will move down. This relation is
quantified by the price impact function
\cite{Hasbrouck91,HaLoMc92,kempf99,pler2002,Rosenow02,EvLy02,LiFaMa03,Ga+03,potbou2002,Hop02,Bou+03},
which describes stock price changes as a conditional expectation value
of the order imbalance. The order imbalance is measured as the
difference between the number of shares bought and the number of
shares sold in a given time interval.  

Gabaix et al. \cite{Ga+03} have
suggested that large price changes are due to large order imbalances.
Starting from the distribution of trading volumes and a fit to the
average price impact function, they suggest an explanation for the
empirical power law distribution of stock price changes. This approach
was criticized in \cite{FaLi03}, because the test presented in
\cite{Ga+03} lacks power in the presence of correlations in the order
flow and because the functional form used to describe the price impact
of large orders  seems to vary for different stock markets.
Instead, the authors of \cite{Fa+03} conclude from an event based
analysis that large price changes are due to the granularity of the
order book, which gives rise to a time varying liquidity.

Here, we present an empirical study of extreme stock price changes
within time intervals of a length $\Delta t = 5 {\rm min}$.  We
analyze one year of data for the 44 most frequently traded NASDAQ
stocks. These data are contained in the Trades and Quotes (TAQ) data
base published by the New York Stock Exchange.  In addition, we
analyze one year of order book data from the Island ECN for the
ten most frequently traded companies \cite{ticker}.  For both data
bases, we find little evidence that price changes larger than five
standard deviations are explained by the order imbalance.  For the
order book data, we are able to reconstruct the price impact function
for time intervals with large price changes and find that price
changes are quantitatively explained by unusually large slopes of the
price impact function.

The TAQ data base contains information about transaction data like the
number of shares and transaction price as well as information about
quotes, i.e.  the lowest sell offer (ask price $S_{\rm ask}(t)$) and
the highest buy offer (bid price $S_{\rm bid}(t)$). The stock price
change or return in a time interval $\Delta t$ is defined as
%
%******************* return definition ******************************
\begin{equation}
G(t) = \ln S_{\rm M}(t+\Delta t) - \ln S_{\rm M}(t),
\end{equation}
%********************************************************************
%
where the midquote price $S_{\rm M}(t)={1 \over 2} (S_{\rm bid}(t) +
S_{\rm ask}(t))$ is the arithmetic mean of bid and ask price.  The
order imbalance $Q$ in a time interval is the sum of all signed market
orders executed between $t$ and $t+\Delta t$.  For the TAQ data, the
sign of a transaction is determined by the Lee and Ready algorithm,
which compares the transaction price to the midquote price. The sign
is positive for buy orders (transaction price larger than midquote
price) and negative for sell orders (transaction price smaller than
midquote price). For the order book data, the data base contains
information about the direction of a trade.  We choose $\Delta t = 5
{\rm min}$. Returns $G$ are normalized by their standard deviation
$\sigma_G$, volumes $Q$ by $\sigma_Q=\left \langle | Q - \langle Q
  \rangle| \right \rangle$ as their cumulative distribution function
follows a power law with exponent close to two. The variance for data
with such a distribution is not well defined.\\

%
%**************************** figure (virtual) price impact *************
\begin{figure}
  \centerline{ \epsfig{file=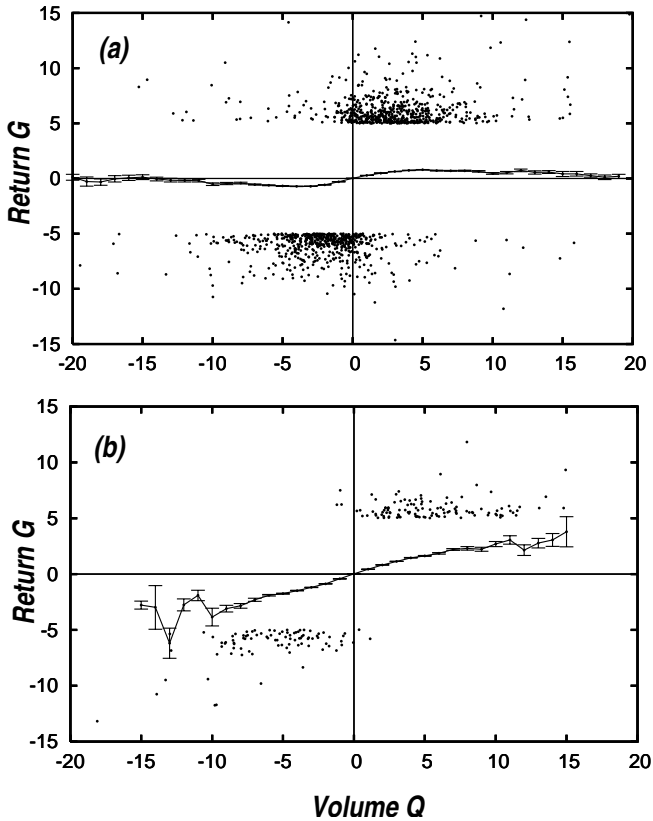,width=8.8cm}} \caption{(a)
    Average price impact function for the 44 most frequently traded
    NASDAQ stocks in the year 1997 with standard deviation of the
    mean. Price changes larger than five standard deviations cluster
    in the region of small volume imbalance, all of them are clearly
    outside the error bars. (b) Same as (a) but for 2002 data from the
    Island ECN order book for the ten most frequently traded stocks.}
  \label{taqscatter.fig}
\end{figure}
%***********************************************************************
%

{\bf Average price impact and large events:}
The relation between price change and order imbalance is described
by the price impact function
%
%*********************** price impact definition ******************
\begin{equation}
I_{\rm market}(Q) = \langle G_{\Delta t}(t) \rangle_{Q} \ \ .
\end{equation}
%*****************************************************************
%
It describes the average price change $G$ caused by an order imbalance
$Q$  \cite{hidden}in the same time interval.

We ask whether the average price impact function $I_{\rm market}(Q)$
is able to describe extremely strong price changes $G > 5 \;
\sigma_G$. We determined all time intervals with price changes larger
than five standard deviations and checked carefully that these large
price changes are not due to errors in the data set but correspond to
"real" events. While the order book data seem to be free of errors,
some errors are contained in the TAQ data.  We have filtered the raw
TAQ data against recording errors and apparent price changes due to
the combination of data from different ECNs (electronic communications
networks).  We have used the algorithm of Chordia et al.
\cite{chordia2001}, which discards all trades where the difference
between trade price and midquote price is larger than 4 times the
spread. The spread is defined as $S_{\rm ask} - S_{\rm bid}$.  In
addition, we have checked visually the return and trading volume time
series surrounding the large price change on a tick by tick basis and
have found no evidence for data errors after applying the filtering
algorithm. The data filtering removes about one percent of all
transactions and has a significant effect on the exponent of the
cumulative distribution function $P(G > X) \sim x^{- \alpha}$.  For
the raw data without any filtering, we find $\alpha = 2.1$, after
applying the filter we find $\alpha = 3.9$ by fitting a straight line
in a double logarithmic diagram. We note that the filtering algorithm
\cite{chordia2001} is very restrictive in the sense that it discards
quite a few events where the TAQ data set reports erratic and strong
oscillations (of several $\sigma_G$) of the price which are probably
due to the combination of data from different ECNs. While the price
has already reached its new "true" value in the leading ECN, there may
still be limit orders at the old price in some smaller ECNs which are
exploited by arbitrage traders.  While these oscillations are "true"
price changes in the sense that they are not due to recording errors,
they are an artifact of the trading system and were not included in
our analysis.

Figure~\ref{taqscatter.fig} shows both the price impact function and
those events with price changes larger than 5 standard deviations
$\sigma_G$. We find 1198 such events for the TAQ data base and 210 for
the Island ECN data.  The large events cluster at quite small values
of $Q$ where the price impact function is significantly below
$G=5\sigma_G$.  Surprisingly, for some of these events not even the
sign of $Q$ and $G$ agree. We believe that this disagreement is mainly
caused by the inaccuracy of the Lee and Ready algorithm, but the
analysis of order book data reveals the existence of such situations
as well. We note that even for large volume imbalances the average
price impact function is several standard deviations (measured by the
statistical error of the mean) below five $\sigma_G$ for the TAQ data.

%
%**************************** figure correlation g and g_pred **********
\begin{figure}
  \centerline{ \epsfig{file=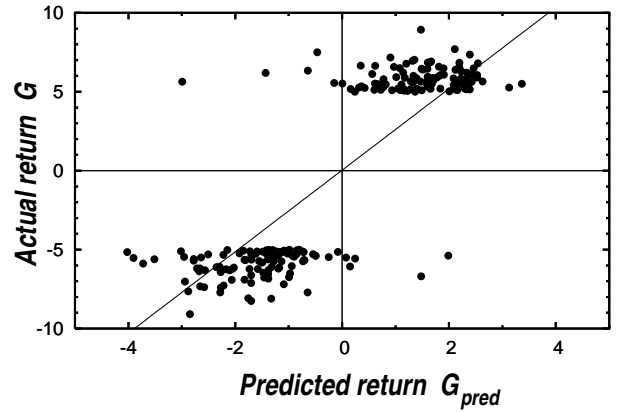,width=8cm}} \caption{Relation
    between the actual return $G$ and the predicted return $G_{\rm
      pred}$ as calculated from the average price impact function
    $I_{\rm market}(Q)$ and the actual order volume.}
\label{g_g_pred.fig}
\end{figure}
%***********************************************************************
%

In Figure~\ref{g_g_pred.fig}, the actual returns $G(t)$ are plotted
against the predicted returns $G_{\rm pred}(t)= I_{\rm market}(Q(t))$
for the order book data. A linear fit to the data points has a slope
of 2.58 indicating that the predicted returns are considerably smaller
than the average one.  The linear fit has an correlation coefficient
$R^2=0.72$ and is not convincing visually. We conclude that the main
cause for large returns is not a large imbalance between buy and sell
orders but some other effect.

%
%**************************** figure (virtual) price impact *************
\begin{figure}
\centerline{ \epsfig{file=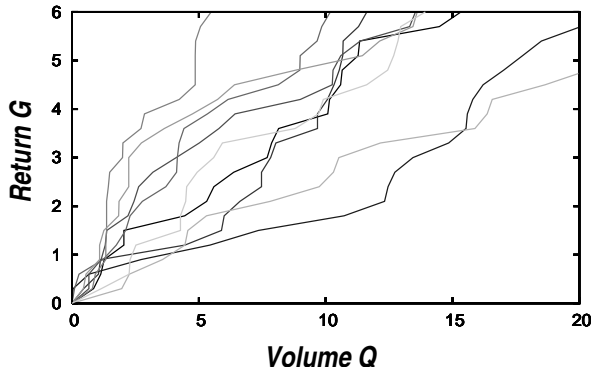,width=8cm,height=5cm}}
\caption{Price change as a function of buy or sell volume for ten of
the largest price changes in the Island ECN data.}
\label{samplepifs.fig}
\end{figure}
%***********************************************************************
%

{\bf Time varying price impact:} As the average price impact function
does not provide for a satisfactory explanation of large returns, we
study the time dependent price impact. In a modern electronic market
place, market orders are matched with limit orders stored in the order
book.  A buy limit order indicates that a trader is willing to buy a
specified number of shares at a given or lower price, while a sell
limit order signals that a trader wants to buy a certain number of
shares at a given or higher price. The buy limit order with the
highest price determines the bid price, the sell limit order with the
lowest price the ask price. The price change due to a given market
order is determined by the limit orders stored in the order book. If a
trader places a buy market order with volume $\Delta Q$, it executes
as many limit orders as necessary to fill that volume.  In this way,
the order book determines the price change due to a single market
order. We describe the order book by a density function $\rho_{\rm
  book}(\gamma, t)$, where the coordinate
%
%****************************  definition of return ********************
\begin{equation}
\gamma = \left\{ \begin{array}{ccc}
(\ln(S_{\rm{limit}}) - \ln(S_{\rm{bid}})) & \mbox{limit buy order}\\
(\ln(S_{\rm{limit}}) - \ln(S_{\rm{ask}})) & \mbox{limit sell
order}
\end{array}\right.   \  .
\label{return.eq}
\end{equation}
%*************************************************************************
%
describes the position in the order book. In our analysis, the
orderbook density is defined on a discrete grid with spacing $0.3 \;
\sigma_G$. A small market buy order with volume $\Delta Q$ causes a
return $\Delta G$. For such an order, volume and return are related
via $\Delta Q = \rho(0+,t) \Delta G$.  For a larger order volume the
relation is
%
%************************ inverse price impact function ******************
\begin{equation}
Q = \int_0^{G} \rho(\gamma,t) \; d\gamma \ \ .
\label{inversepif.eq}
\end{equation}
%*************************************************************************
%
The return $G$ defined by Eq.~\ref{inversepif.eq} is denoted as the
instantaneous or virtual price impact.  From this relation one sees
that the same order volume $Q$ can be related to quite different
returns $G$ depending on the value of $\rho(\gamma,t)$. In the
following, we will argue that it is this time dependence of the order
book which is responsible for the occurrence of large price changes.
We note that from the order book one obtains only information about
the price change as a function of buy or sell volume. Order book
information can be related to time aggregated signed order volumes
only under the assumption i) that the order book is symmetric around
the midquote price and that ii) nonlinearities can be neglected. Both
assumptions are generally not satisfied. For this reason, we will
consider either the buy or the sell volume in a given five minute
interval, depending on the direction of the return in that interval.

%
%**************************** figure (virtual) price impact *************
\begin{figure}
\centerline{ \epsfig{file=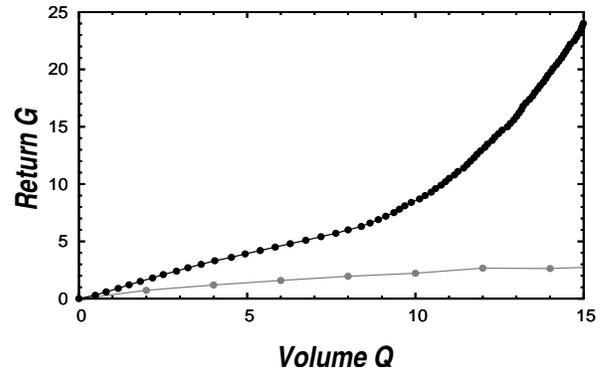,width=8cm,height=5cm}}
\caption{Average price change as a function of buy or sell volume for
  all price changes larger than $5 \sigma_G$ in the Island ECN data.
  The average price change for all transactions is much smaller than
  that for the extreme events. }
\label{averagepif.fig}
\end{figure}
%***********************************************************************

When studying the price impact of the order flow in a given time
interval, it is not sufficient to invoke the order book density
$\rho_{\rm book}(\gamma,t)$ at one instant of time. In addition, one has to
consider changes in the order book which occur in this time interval.
In \cite{WeRo03} it was shown that the virtual price impact of a given
order volume is roughly four times stronger than the actual price
impact. This difference is due to additional limit orders placed in
reaction to a price change. From this example one sees that the
inclusion of dynamical effects is crucial for calculating the correct
price impact.

In order to calculate the density of additional limit orders arriving
in a given time interval, we fix the reference frame by the midquote
price in the beginning of the interval. The density of incoming
limit orders is denoted by $\rho_{\rm flow}(\gamma,t)$, and the
total order density is given by
%
%****************************  definition of order density ***************
\begin{equation}
\rho(\gamma,t) = \rho_{\rm book}(\gamma,t) + \rho_{\rm flow}(\gamma,t,\Delta t) \ \ .
\label{density.eq}
\end{equation}
%*************************************************************************
%
From $\rho(\gamma,t)$ we calculate a price impact function $I_{\rm
  actual}(Q)$ by inverting the relation Eq.~\ref{inversepif.eq}.  The
sell order side of this function for ten events with price changes
larger than $5\sigma_G$ is shown in Figure \ref{samplepifs.fig}.  In
Figure~\ref{averagepif.fig}, the average over all such events is
compared to the average price impact function $I_{\rm market}(Q)$. One
sees that the slope of $I_{\rm actual}(Q)$ is much steeper than the
slope of $I_{\rm market}(Q)$.

%
%**************************** figure (virtual) price impact *************
\begin{figure}
\centerline{ \epsfig{file=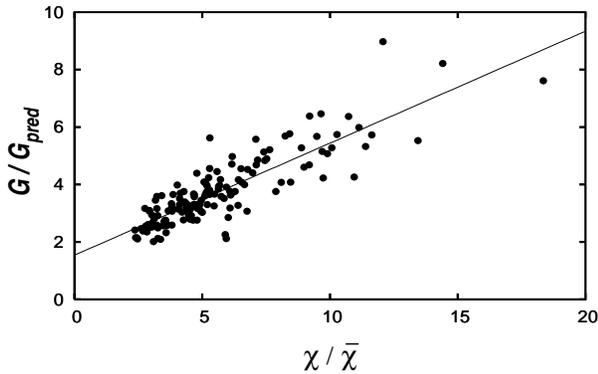,width=8cm,height=5cm}}
\caption{Ratio of actual price change to  predicted price change plotted
against the slope of the actual price impact function 
normalized by the slope of the average price impact function.
The data points cluster in the vicinity of a linear fit. }
\label{scatterplot.fig}
\end{figure}
%***********************************************************************

The fact that the price impact function for large events has a steeper
slope than the average price impact function implies that in  time
intervals with large price movements there are less limit orders
available than on average. Hence,  the slope of the actual price impact
function provides a measurement of the market liquidity.

Some curves displayed in Figure~\ref{samplepifs.fig} show marked
nonlinearities. The exponents found from power law fits vary between
0.15 and 2.35 with a mean of 1.32 and a standard deviation of 0.41.
However, the average of the $I_{\rm actual}$ for all large events (see
Figure~\ref{averagepif.fig}) is approximately linear, a power law fit
yields an exponent of 1.03.  As a measure for the strength of price
impact, we define a susceptibility $\chi(t)$ for the actual price
impact function for a given time interval by a linear fit up to a
price change of $G=5\sigma_G$.  Using this definition, we look for an
explanation of extreme price changes.  We compare the ratio of the
actual price change $G(t)$ and the predicted price change
%
%******************************* predicted price change  *****************
\begin{equation}
G_{\rm pred}(t) = I_{\rm market}(Q(t)) \ \ .
\end{equation}
%*************************************************************************
%
to the ratio of actual slope $\chi(t)$ and slope $\chi_{\rm market}$
of the average price impact function $I_{\rm market}$.  As explained
above, a price impact calculated from the order book can only be
defined for either buy or sell volume. For this reason, we have
recalculated $I_{\rm market}$ by averaging with respect to either the
sell or the buy volume, depending on the sign of the price change.
This recalculated $I_{\rm market}$ is quite similar to the original
one.  To calculate $G_{\rm pred}(t)$, we use the buy volume for
positive returns and the sell volume for negative ones.

In Figure \ref{scatterplot.fig} the ratio of $G_{\rm pred}$ and $G$ is
plotted against the susceptibility $\chi/ \chi_{\rm market}$ for all
events with $|G|>5\sigma_G$. We see that the data points cluster in
the vicinity of a straight line fit with an $R^2= 0.74$. We conclude
that the time dependent slope of the price impact function has a large
explanatory power for the occurrence of extreme price changes.

In summary, we have studied two alternative approaches to explain
large stock price changes: large fluctuations in trading volume and
time changing liquidity for two different data sets.  We find little
evidence that extreme stock price changes are caused by large trading
volume. For order book data, we are able to reconstruct the price
impact for the time intervals with large returns. We find that the
slope of this price impact strongly correlates with the ratio of
observed return and the return predicted from the trading volume and
the average price impact function.


\begin{thebibliography}{99}
\bibitem{hidden}  We do not include market orders executing "hidden"
limit orders in the definition of $Q(t)$ as we want to compare with the
order book containing "visible" orders only.
\end{thebibliography}
\end{document}